\RequirePackage{lineno}
\RequirePackage{ulem}

\documentclass[%
 aps,
 jmp,%
amsmath,amssymb,
preprint,
endfloats*,%
showpacs,
]{revtex4-1}

\usepackage{graphicx}
\usepackage{dcolumn}
\usepackage{bm}

\begin{document}
\preprint{AIP/123-QED}

\title[]{New structural model for GeO$_2$/Ge interface: A first-principles study}

\author{Shoichiro Saito}
\email{saito@cp.prec.eng.osaka-u.ac.jp}
\author{Tomoya Ono}%
\affiliation{ 
Graduate School of Engineering, Osaka University
}%

\date{\today}

\begin{abstract}
First-principles modeling of a GeO$_2$/Ge(001) interface reveals that sixfold GeO$_2$, which is derived from cristobalite and is different from rutile, dramatically reduces the lattice mismatch at the interface and is much more stable than the conventional fourfold interface. Since the grain boundary between fourfold and sixfold GeO$_2$ is unstable, sixfold GeO$_2$ forms a large grain at the interface. On the contrary, a comparative study with SiO$_2$ demonstrates that SiO$_2$ maintains a fourfold structure. The sixfold GeO$_2$/Ge interface is shown to be a consequence of the ground-state phase of GeO$_2$. In addition, the electronic structure calculation reveals that sixfold GeO$_2$ at the interface shifts the valence band maximum far from the interface toward the conduction band. 
\end{abstract}


\maketitle
\linenumbers 

With the continued scaling of Si metal-oxide-semiconductor (MOS) devices, it is becoming increasingly difficult to enhance device performance; therefore, new channel materials have been explored. Ge is considered as one of the best channel materials for obtaining high performance MOS devices due to its higher carrier mobility. The narrower band gap of Ge is also attractive since lower operation voltage lowers energy consumption. To fabricate high performance devices with a Ge channel, one of the most crucial challenges is fabricating an insulator with superior interface properties since the termination of the surface states is important for device reliability. So far, various insulators have been examined, for example, GeO$_2$ \cite{matsubara,hosoi}, Ge$_3$N$_4$ \cite{ge3n4}, and high-\textit{k} oxides \cite{high-k1,high-k2,summary}. Among these insulators, GeO$_2$ is the most fundamental and important, similar to SiO$_2$ in Si MOS technology, because it exists even in high-\textit{k} oxide/Ge interfaces. 
A considerable number of studies have been conducted on the process of developing high quality GeO$_2$/Ge interfaces, and several groups have reported that GeO$_2$/Ge interfaces, fabricated by conventional dry oxidation, have a low interface trap density (mid 10$^{10}$--10$^{11}$ cm$^{-2}$eV$^{-1}$) \cite{matsubara,hosoi}. 
Houssa \textit{et al.} simulated the density of Ge dangling bonds at the GeO$_2$/Ge interface as a function of the oxidation temperature, by combining viscoelastic data of GeO$_2$ and the modified Maxwell's model, and claimed that the density of Ge dangling bonds is less than that of Si dangling bonds at the SiO$_2$/Si interface \cite{houssa}. Their results are in good agreement with other reported results \cite{matsubara,hosoi}. 

Interface stress between a semiconductor and an oxide is considered as one of the origins of interface defects. Kageshima and Shiraishi predicted that Si atoms are emitted from the interface to release stress induced by the lattice-constant mismatch between SiO$_2$ and Si, although the dangling bonds remain after Si atom emission \cite{prl}. 
On the other hand, we calculated the probability of Ge atom emission from a GeO$_2$/Ge interface by using the interface model proposed by Kageshima and Shiraishi and found that hardly any Ge atoms are emitted from the GeO$_2$/Ge interface \cite{apl}. We concluded that the flexibility of the O-Ge-O bonds contributes to the relaxation of interface stress, resulting in a GeO$_2$/Ge interface that is superior to the SiO$_2$/Si one. 
Watanabe \textit{et al.} claimed, using the classical molecular-dynamics simulation, that the narrow equilibrium Ge-O-Ge bond angles contribute to the reduction in compressive stress in GeO$_2$ films as well as flexible O-Ge-O bonds \cite{ecs}. Although a considerable number of first-principles studies on the GeO$_2$/Ge interface have been conducted based on the calculations on SiO$_2$/Si interfaces \cite{houssa,yang,pas,pas2,kummel,pant}, the atomic and electronic structures of the GeO$_2$/Ge interface have not been identified experimentally because GeO$_2$ is both water-soluble and thermally unstable at elevated temperatures. 

We propose a complete ordered GeO$_2$/Ge(001) interface structure with minimum lattice mismatch. Our interface model consists of sixfold GeO$_2$, which is derived from cristobalite and is different from rutile, on a Ge(001) substrate. The sixfold GeO$_2$/Ge interface is more stable by 1.92 eV than the conventional fourfold GeO$_2$/Ge interface. The lattice mismatch between sixfold GeO$_2$ and Ge ($\sim$5$\%$) is much smaller than that between fourfold GeO$_2$ and Ge ($\sim$17$\%$). By examining a mixed fourfold and sixfold GeO$_2$/Ge interface model, we find that sixfold GeO$_2$ exists as a large grain at the GeO$_2$/Ge interface. On the contrary, SiO$_2$ at the SiO$_2$/Si interface maintains a fourfold structure. 
A comparative study of the electronic structures of the sixfold and fourfold GeO$_2$/Ge interfaces shows that the valence band maximum (VBM) far from the interface varies due to the existence of sixfold GeO$_2$ at the interface. 

\begin{figure}
\includegraphics{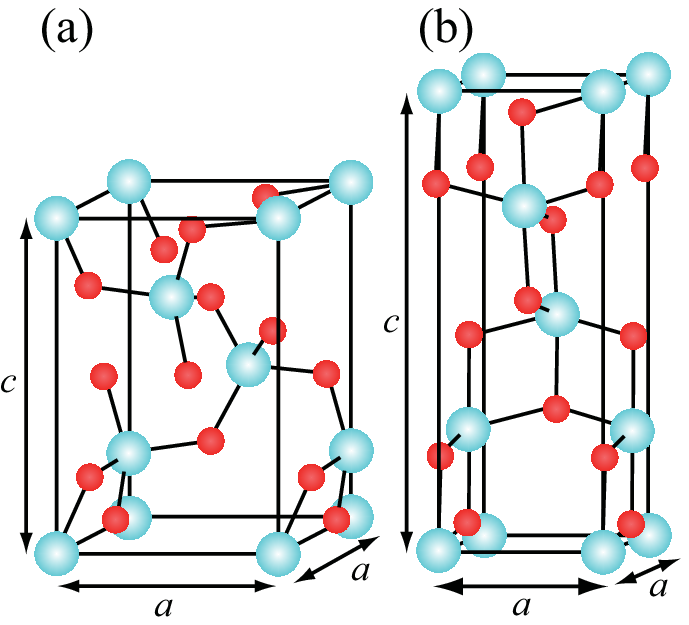}
\caption{\label{fig:fig1} (Color online) Unit cells of (a) fourfold and (b) sixfold \textit{c}-GeO$_2$. The solid cube represents the unit-cell volume, and the blue (light) and red (dark) circles are Ge and O atoms, respectively. }
\end{figure}

Our first-principles calculation method is based on the real-space finite-difference approach \cite{rs1,rs2,rs3}, which enables us to determine a self-consistent electronic ground state with a high degree of accuracy using a timesaving double-grid technique \cite{rs2,rs3}. The norm-conserving pseudopotentials \cite{ps} of Troullier and Martins \cite{tm} are used to describe the electron-ion interaction and are transformed into the computationally efficient Kleinman-Bylander separable form \cite{kb}, using the $s$ and $p$ components as nonlocal components for H, O, and Si, and the $s$, $p$, and $d$ components as nonlocal components for Ge. Exchange and correlation effects are treated using the local density approximation \cite{lda}. The coarse grid spacing of $\sim$0.13 \AA, which corresponds to the plane wave cutoff energy of $\sim$112 Ry, is used for all of our calculations. 
We first examine the atomic structures of GeO$_2$ and SiO$_2$ bulks in the cristobalite phases (\textit{c}-GeO$_2$ and \textit{c}-SiO$_2$) under pressure along the $a$-axis because these structures correspond to the directions parallel to the interface when the oxides are piled on the (001) surface. 
The \textit{c}-GeO$_2$ (\textit{c}-SiO$_2$) structure is tetragonal with four GeO$_2$ (SiO$_2$) molecules per unit cell, and 4 $\times$ 4 $\times$ 3 \textit{k}-point grids in the Brillouin zone are taken into account. 
The \textit{c}-GeO$_2$ structure at the equilibrium point is shown in Fig.~\ref{fig:fig1}(a). 
The equilibrium lattice parameters of \textit{c}-GeO$_2$ and \textit{c}-SiO$_2$, which are obtained from first-principles calculation, are listed in Table~\ref{tbl:lattice}. We then compress \textit{c}-GeO$_2$ (\textit{c}-SiO$_2$) along the $a$-axis by 5-25 (5-35)$\%$ from the equilibrium lattice constants and optimize the length of the $c$-axis in increments of 1$\%$ to determine the energy minima. 
We relax all the atoms until all the force components drop below 0.05 eV/\AA. 

\begin{figure}
\includegraphics{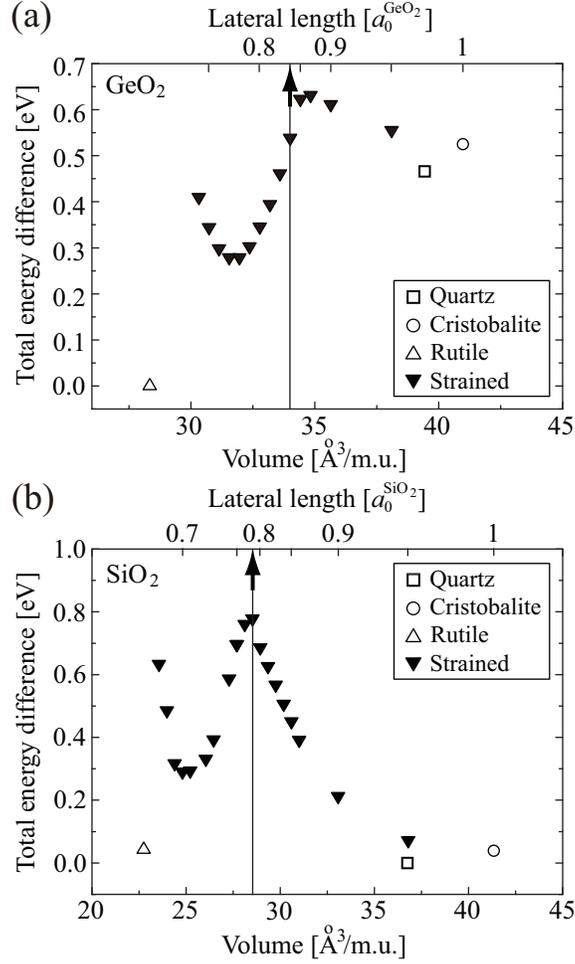}
\caption{\label{fig:fig2} Total energy per molecular unit (m.u.) as a function of volume for (a) \textit{c}-GeO$_2$ and (b) SiO$_2$. Energy minima of other phases [quartz ($\square$), cristobalite ($\bigcirc$), and rutile ($\triangle$)] are also shown for comparison. The zero on the energy scale is rutile for GeO$_2$ and quartz for SiO$_2$. The upper horizontal axes correspond to the lateral lengths of Ge and Si(001)-($1 \times 1$) surfaces in $a_0^{\rm{GeO_2}}$ and $a_0^{\rm{SiO_2}}$, where $a_0^{\rm{GeO_2}}$ and $a_0^{\rm{SiO_2}}$ represent the lengths of the $a$-axes of \textit{c}-GeO$_2$ and \textit{c}-SiO$_2$ at the equilibrium points, respectively. }
\end{figure}

Figures~\ref{fig:fig2}(a) and ~\ref{fig:fig2}(b) show the total energies of \textit{c}-GeO$_2$ and \textit{c}-SiO$_2$ per molecular unit as a function of volume. The energy minima of other phases [quartz ($\square$), cristobalite ($\bigcirc$), and rutile ($\triangle$)] without any constraints are also depicted for comparison. The zero on the energy scale is the rutile structure of GeO$_2$ and the quartz structure of SiO$_2$. 
In this figure, the \textit{c}-GeO$_2$ at about 0.78$a_0^{\rm{GeO_2}}$  shows a local minimum, where $a_0^{\rm{GeO_2}}$  represents the length of the $a$-axis of \textit{c}-GeO$_2$ at the equilibrium point (4.818 \AA). 
The atomic structure of the strained \textit{c}-GeO$_2$ at the local minimum is shown in Fig.~\ref{fig:fig1}(b). 
The \textit{c}-GeO$_2$ under a certain pressure transforms into a sixfold structure, which is distinct from the rutile phase, by rotating oxygen atoms around the Ge atoms. The energy minimum of sixfold GeO$_2$ is lower than that of fourfold GeO$_2$ since the zero-temperature phase of GeO$_2$ has a sixfold rutile structure. 
The critical point between the fourfold and sixfold structures is about 0.85$a_0^{\rm{GeO_2}}$. The arrow on the upper horizontal axis corresponds to the lateral length of the Ge(001)-($1 \times 1$) surface. It should be noted that GeO$_2$ forms a sixfold structure when the length of the $a$-axis is equal to that of the ($1 \times 1$) surface, while the \textit{c}-SiO$_2$ still maintains a fourfold structure. 

\begin{figure}
\includegraphics{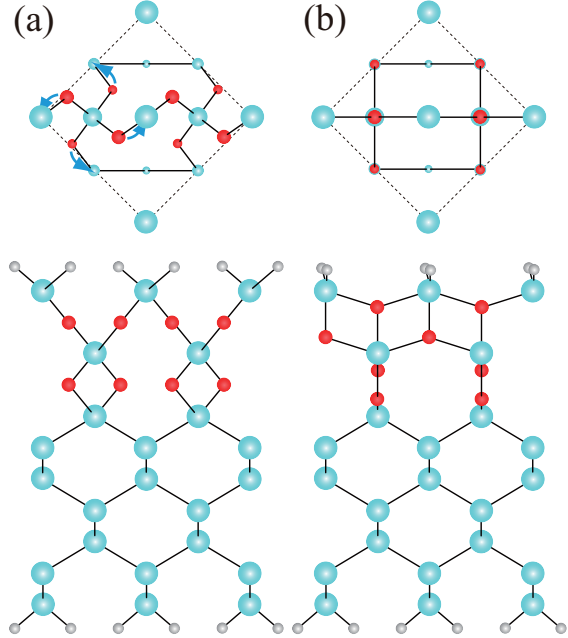}
\caption{\label{fig:fig3} (Color online) Top views and side views of (a) fourfold and (b) sixfold GeO$_2$/Ge(001) interfaces. The blue (light), red (dark), and grey (light small) circles are Ge, O, and H atoms, respectively. The dotted square in the top views represents a Ge(001)-($\sqrt{2} \times \sqrt{2}$) surface unit and the arrows indicate rotational directions to transform into sixfold structures.}
\end{figure}

We next compare the energetic stability of the sixfold GeO$_2$/Ge(001) interface with the fourfold one since the lattice mismatch between the sixfold GeO$_2$ and Ge(001) surfaces is small ($\sim$5$\%$). Figures~\ref{fig:fig3}(a) and ~\ref{fig:fig3}(b) show the fourfold and sixfold GeO$_2$/Ge interfaces, respectively. The fourfold oxide transforms into a sixfold one by rotating the four oxygen atoms around one Ge atom in a Ge(001)-($1 \times 1$) surface unit. The total energy difference between the fourfold and sixfold SiO$_2$/Si interfaces are also calculated for comparison. The Ge(001)-($\sqrt{2} \times \sqrt{2}$) surface is used for the lateral size of the supercell of the GeO$_2$/Ge interface, and the length of the supercell perpendicular to the surface is 5.5$a_0^{\rm{Ge}}$, where $a_0^{\rm{Ge}}$ is the optimized lattice constant of the Ge bulk (5.578 \AA). The model of the interface includes seven Ge atomic and two GeO$_2$ molecular layers, and both sides of the surface are simply terminated with H atoms. Eight \textit{k}-points in the 1 $\times$ 1 lateral unit cell are used for the Brillouin zone sampling. All the atoms, except the Ge atoms in the bottom-most layer and the H atoms terminating their dangling bonds, are relaxed. The other computational details are the same as those used in the bulk calculation. 
We found that the sixfold GeO$_2$/Ge interface is more stable by 1.92 eV than the fourfold one because the interface stress between GeO$_2$ and Ge is released by the phase transition into the dense sixfold structure. On the other hand, the fourfold \textit{c}-SiO$_2$/Si interface model is preferable by 1.02 eV compared with the sixfold one. 
Kageshima and Shiraishi reported that Si atoms at the \textit{c}-SiO$_2$/Si interface are emitted to release interface stress, resulting in a quartz-SiO$_2$/Si interface \cite{prl}. On the other hand, hardly any Ge atoms at the \textit{c}-GeO$_2$/Ge interface are emitted \cite{apl}. Our results indicate that the sixfold structure contributes to the release of interface stress due to the lattice mismatch between the \textit{c}-GeO$_2$ and Ge(001) surfaces. 

\begin{figure}
\includegraphics{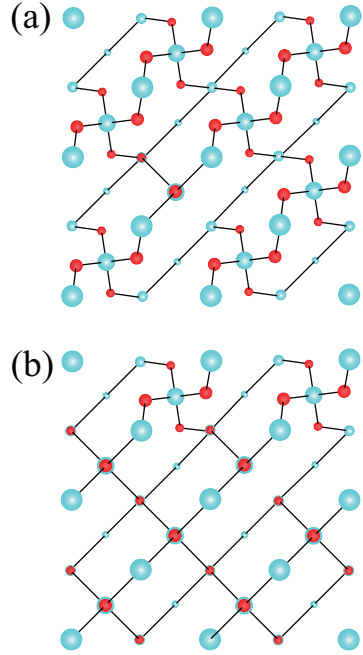}
\caption{\label{fig:fig4} (Color online) Top views of mixed fourfold and sixfold structures. (a) 12.5$\%$ and (b) 62.5$\%$ of Ge(001)-($1 \times 1$) surface units are composed of the sixfold structures. The blue (light) and red (dark) circles are Ge and O atoms, respectively. }
\end{figure}

Since the lateral length of the Ge(001)-($1 \times 1$) surface is longer than that of the sixfold GeO$_2$ surface but shorter than that of the fourfold \textit{c}-GeO$_2$ surface in Fig.~\ref{fig:fig2}(a), there is a possibility that \textit{c}-GeO$_2$ on the Ge(001) surface is composed of a mixed fourfold and sixfold structure. 
We examine the total energies of the supercell doubling of the Ge(001)-($\sqrt{2} \times \sqrt{2}$) surface unit in the two directions, i.e., the supercell contains eight Ge(001)-($1 \times 1$) units. We respectively replace one and five neighboring ($1 \times 1$) Ge surface units so that 12.5$\%$ and 62.5$\%$ of the Ge(001)-($1 \times 1$) units are composed of the sixfold structures [Figs.~\ref{fig:fig4}(a) and ~\ref{fig:fig4}(b)]. 
The computational procedures are the same as mentioned above. 
The calculated total energy differences are summarized in Table~\ref{tbl:ratio} with respect to the ratio of the sixfold coordination. 
The fully sixfold GeO$_2$/Ge interface is the most stable, and the mixed interface with the 12.5$\%$ sixfold structure is even more unstable than the fully fourfold GeO$_2$/Ge interface. 
The instabilities of the mixed interfaces are attributed to the grain boundaries; the $c$-axis of the fourfold oxidized region is more than 5$\%$ longer than that of the sixfold one. 
This result implies that the sixfold oxidized region exists as a large grain at the GeO$_2$/Ge interface. 

\begin{figure}
\includegraphics{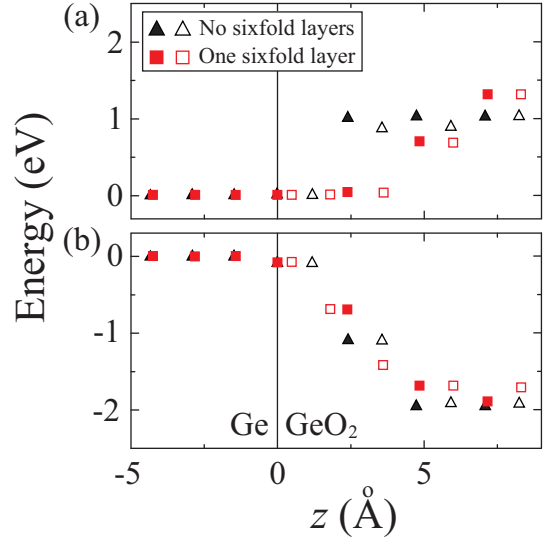}
\caption{\label{fig:fig5} (Color online) Variations of (a) CBM and (b) VBM along orthogonal direction to interface obtained on each atom. Triangles and squares are CBM and VBM of GeO$_2$/Ge interfaces with zero and one sixfold GeO$_2$ layers, respectively. Filled (open) symbols represent CBM and VBM on Ge (O) atoms. Solid line indicates boundary between GeO$_2$ and Ge(001) substrate. } 
\end{figure}

Finally, we investigate the effect of the sixfold structure on the variations of the conduction band minimum (CBM) and VBM along the normal direction to the interface. To suppress the effect of quantum confinement due to the limitation of substrate thickness, a 12-atomic layer of the Ge(001) substrate is used. The four GeO$_2$ molecular layers are piled on the Ge(001) substrate, and other computational details are the same. 
It is believed that GeO$_2$ is composed of an amorphous structure. Tamura \textit{et al.} found that the band gap of crystalline GeO$_2$ is compatible with that of amorphous GeO$_2$ by using the first-principles calculation \cite{tamura}. Therefore, we calculate the electronic structure with and without a sixfold GeO$_2$ layer inserted between the crystalline fourfold GeO$_2$ layer and Ge(001) substrate. 
The grain boundary between the sixfold and fourfold structures parallel to the Ge(001) substrate is stable, although fivefold Ge atoms exist between the sixfold and fourfold boundaries. 
Figures~\ref{fig:fig5}(a) and ~\ref{fig:fig5}(b) show the evolution of the CBM and VBM along a direction orthogonal to the interface plane, respectively. CBM and VBM variations are subtracted from the local density of states, calculated by integrating them on the plane parallel to the interface based on $\rho(z,E)=\int |\psi(\mbox{\boldmath$r$},E)|^2 d\mbox{\boldmath$r$}_{||}$, with a contour of 7.94 $\times 10^{-5}$~$e$/eV/\AA$^3$. 
With the fourfold GeO$_2$/Ge interface, the VBM is almost complete at about 5 \AA~deep from the interface, which is similar than that with the SiO$_2$/Si interface \cite{yamasaki}, while the CBM saturates within 2 \AA. 
Since the band gap of sixfold GeO$_2$ is narrower than that of fourfold GeO$_2$ \cite{comment}, the valence electrons in the Ge substrate penetrate sixfold GeO$_2$ and the interface dipole emerges. Therefore, the existence of the sixfold GeO$_2$ layer shifts the VBM far from the interface toward the conduction band. 

In summary, we proposed a sixfold GeO$_2$/Ge interface, in which the lattice mismatch at the interface is very small ($\sim$5$\%$) and which is energetically much more stable than fourfold GeO$_2$/Ge interfaces. It should be noted that the sixfold structure was found to be a large grain at the GeO$_2$/Ge interface after computing the stability of the mixed fourfold and sixfold GeO$_2$/Ge interface. On the other hand, with SiO$_2$, a conventional fourfold structure on the Si(001) substrate is preferable due to the difficulty in rearranging the rigid O-Si-O bonds even in the bulk phase. 
The electronic structure calculation with and without the sixfold GeO$_2$ monolayer at the GeO$_2$/Ge interface reveals that the VBM far from the interface in the ultrathin GeO$_2$ layer ($\sim$10 \AA) depends on the coordination number of GeO$_2$. 
Our results provide new insight into a strong candidate of the atomic and electronic structures of the GeO$_2$/Ge interface. We await experimental verification of our prediction. 

\begin{table}
\caption{\label{tbl:lattice} Lattice constants of \textit{c}-GeO$_2$ and \textit{c}-SiO$_2$ at their equilibrium points. The unit is \AA }
\begin{ruledtabular}
\begin{tabular}{ccc}
Structure & $a$ & $c$\\
\hline
\textit{c}-GeO$_2$ & 4.818 & 7.128 \\
\textit{c}-SiO$_2$ & 4.925 & 6.828 \\
\end{tabular}
\end{ruledtabular}
\end{table}

\begin{table}
\caption{\label{tbl:ratio} Energy difference between fully fourfold \textit{c}-GeO$_2$ structure and various mixing ratios of fourfold and sixfold structures with respect to ratio of sixfold structure. All units are in eV. }
\begin{ruledtabular}
\begin{tabular}{ccccc}
Ratio of sixfold structure & 0$\%$ & 12.5$\%$& 62.5$\%$& 100$\%$\\
\hline
 & 0 & 0.92 & -0.49 & -7.67 \\
\end{tabular}
\end{ruledtabular}
\end{table}

The authors would like to thank Professor Kenji Shiraishi of University of Tsukuba, and Professor Heiji Watanabe, Professor Yoshitada Morikawa, Professor Takayoshi Shimura, and Professor Takuji Hosoi of Osaka University for reading the manuscript and for contributing to our fruitful discussions. 
This research was partially supported by the Strategic Japanese-German Cooperative Program from the Japan Science and Technology Agency and Deutsche Forschungsgemeinschaft, by a Grant-in-Aid for Young Scientists (B) (Grant No. 20710078), and also by a Grant-in-Aid for the Global COE "Center of Excellence for Atomically Controlled Fabrication Technology" from the Ministry of Education, Culture, Sports, Science and Technology, Japan. The numerical calculations were carried out using the computer facilities of the Institute for Solid State Physics at the University of Tokyo, the Center for Computational Sciences at University of Tsukuba, the Research Center for Computational Science at the National Institute of Natural Science, and the Information Synergy Center at Tohoku University.


\begin{thebibliography}{99}
\bibitem{comment}
The band gap of $\sim$2.0 eV ($\sim$3.2 eV)  is estimated for the sixfold (fourfold) GeO$_2$ at the local density approximation level.
\end{thebibliography}
\end{document}